# A New Mathematical Formulation of the Governing Equations for the Chemical Compositional Simulation

B. E. Bekbauov[1], A. Kaltayev[1], A. Berdyshev[2]

[1] Al-Farabi Kazakh National University, Almaty, Kazakhstan
[2] Abai Kazakh National Pedagogical University, Almaty, Kazakhstan
Corresponding author: bakhbergen.bekbauov@kaznu.kz

**Abstract**

It is the purpose of this work to develop new approach for chemical compositional reservoir simulation, which may be regarded as a sequential method. The development process can be roughly divided into the following two stages: (1) development of a new mathematical formulation for the sequential chemical compositional reservoir simulation, (2) implementation of a sequential solution approach for chemical compositional reservoir simulation based on the formulation described in this paper. This paper addresses the first stage of the development process by presenting a new mathematical formulation of the chemical compositional reservoir flow equations for the sequential simulation. The newly developed mathematical formulation is extended from the model formulation used in existing chemical compositional simulators. During the model development process, it was discovered that the currently used chemical compositional model estimates the adsorption effect on the transport of a component reasonably well but it violates the principle of mass conservation. The energy conservation equation in the currently used chemical compositional model does not consider any change in the effective pore size caused by adsorption, which leads to inconsistency between the overall compositional balance equations and the energy conservation equation by violating conservation of energy. With these partial differential equations as governing equations, several simulators have been developed. In this article, we propose a formulation to model the change in pore volume due to adsorption that satisfies the conservation laws for mass and energy, and allows applying a sequential solution approach.

**Keywords:** Chemical compositional flow; Model formulation; Reservoir simulation; Chemical flooding; Enhanced oil recovery

## Introduction

One of the important methods in enhanced oil recovery (EOR) is chemical flooding. Chemical flooding is a general term for injection processes that use special chemical solutions. Micellar, alkaline, and soap-like substances are used to reduce surface tension between oil and water in the reservoir, whereas polymers such as polyacrylamide or polysaccharide are employed to improve sweep efficiency. The chemical solutions are pumped through specially distributed injection wells to mobilize oil left behind after primary or secondary recovery. While chemical flooding in the petroleum industry has a larger scale of oil recovery efficiency than water flooding, it is far more technical, costly, and risky. Model studies can assist in this evaluation. The displacement mechanisms in this type of flooding involve interfacial tension lowering, capillary desaturation, chemical synergetic effects, and mobility control, and its flow and transport model describes such physicochemical phenomena as dispersion, diffusion, adsorption, chemical reactions, and in-situ generation of surfactant from acidic crude oil.

Most multiphase compositional models reported in the open literature [1-9] are limited in their applicability in one way or another (single species, equilibrium mass transfer, and lack of modeling miscibility which occurs during surfactant flooding). The only surfactant enhanced aquifer remediation (SEAR) models reported in the literature are for single phase flow and are those of Wilson, Wilson and Clarke and Abriola *et al.* [10-12] with simplified surfactant phase behavior and properties. None of these models account for the effects of surfactant on interfacial tension, surfactant phase behavior, capillary number, and surfactant adsorption.

A simplified thermal mathematical model of the simultaneous flow of immiscible fluids [13, 14] was applied for simulating mechanical and chemical water control (gel/polymer treatment disproportionate permeability reduction) methods.

The mathematical formulation developed in the scope of this work is extended from the UTCHEM model formulation for use in chemical flooding studies that does not have these common limitations.

Several compositional formulations have been developed and exist in the literature [15-31]. These formulations differ in the selection of the primary variables, the degree of implicitness, and the choice of primary equations. Wong and Aziz [32] provided a comprehensive study regarding the most widely used formulations.

Chemical compositional simulation is a complex task that involves solving several equations simultaneously for all grid blocks representing a petroleum reservoir. The sequential schemes are very suitable for the chemical compositional flow problems that involve many chemical components. Only the IMPEC formulation has been used to date for the chemical compositional reservoir simulation, but there is no apparent reason why the sequential formulation could not be used as well. Because of the explicitness for the solution of compositions, the size of time steps must be restricted to stabilize the overall procedure. In contrast, the sequential technique solves the equations for pressure, compositions and energy implicitly, and relaxes the time step restriction.

In 2007, Chen et al. [33] presented a numerical approach that solves both pressure and compositions implicitly. The sequential approach splits the coupled system of nonlinear governing equations of the model up into individual equations and solves each of these equations separately and implicitly. The system of species conservation equations was solved implicitly for the overall concentration of each component. Though the approach was claimed to be sequential and extended from the IMPEC approach used in UTCHEM model [34], the mathematical formulations for the governing equations did not undergo any change in their model. Since the overall concentration appears only in the accumulation term of the species-conservation equations, it remains unclear how the sequential approach was used to solve these equations implicitly for the overall concentration.

Although the existing sequential formulations for compositional models can be applied to the chemical flooding model, they require significant change in the current algorithm used in chemical compositional simulators.

The basic equations used in chemical compositional model that describe multiphase, multicomponent flow in permeable media are the species-conservation equations, pressure (an overall mass-continuity) equation, and energy conservation equations. Accumulation terms in the species-conservation equations used in existing chemical compositional model account for the reduction in pore volume caused by adsorption. During the process of this research, it was revealed that this commonly used approach estimates the adsorption effect on the transport of a component reasonably well but it does not satisfy the species-conservation equation. The energy conservation equation does not take into account any change in the pore volume due to adsorption at all. With these as governing equations, several simulators have been developed during recent years to model chemical compositional phenomena in petroleum reservoirs [33, 34, and 35].

In the present work we introduce a new approach to model the reduction in pore volume due to adsorption that satisfies the continuity equation. In certain situations, such as significant change in the effective pore size due to adsorption, these enhancements are essential to properly model the physical phenomena occurring in petroleum reservoirs. The formulation suggested in this work does not require any overall change in the currently used algorithm and allows us to apply a sequential solution approach for chemical compositional reservoir simulation.

The novelty of the research work consists in the development of a new mathematical formulation of the mass conservation, pressure and energy equations for the sequential chemical compositional simulation. The simplicity of the sequential solution algorithm is believed to be a new contribution as well.

## Mathematical Model Formulation

Consider a bulk volume $V_b$ at some point within a porous medium domain. Let us assume that this representative elementary volume is made up of $n_p + 1$ phases ($n_p$ fluid phases and a solid phase consisting of rock grains or soil) with $n_c$ chemical species. Conceivably, at least, each species can exist in any phase and can transfer between phases via evaporation, condensation, dissolution, adsorption and so forth.

In our model formalism, each pair $(i, \alpha)$, with $i$ chosen from the species indices and $\alpha$ chosen from the phases, is a constituent. Each constituent $(i, \alpha)$ has its own intrinsic mass density $\rho_{i\alpha}$, measured as mass of $i$ per unit volume of phase $\alpha$, and its own average velocity $\vec{u}_{i\alpha}$. Each phase $\alpha$ has its own volume fraction $\phi_\alpha$. The volume fraction of phase $\alpha$, $\phi_\alpha$, is the volume of phase $\alpha$ divided by the bulk volume $V_b$.

If the index $n_c$ represents the number of species and the index $n_p + 1$ represents the number of phases including the solid phase, then in terms of the above defined mechanical variables the mass balance for each constituent $(i, \alpha)$ is

$$\frac{\partial}{\partial t}(\phi_\alpha \rho_{i\alpha}) + \nabla \cdot (\phi_\alpha \rho_{i\alpha} \vec{u}_{i\alpha}) = R_{i\alpha} + r_{mi\alpha} + q_{i\alpha}, \qquad i = 1, \ldots, n_c;\ \alpha = 1, \ldots, n_p + 1. \qquad (1)$$

From left to right in Eq. (1), the terms are now the accumulation, transport, and source terms, the last consisting of three types.

The source term $R_{i\alpha}$ accounts for the rate of mass generation ($R_{i\alpha} > 0$) and consumption ($R_{i\alpha} < 0$) of component $i$ in phase $\alpha$, either through chemical or biological reactions. The volume fractions of each phase are handled through $R_{i\alpha} = \phi_\alpha r_{i\alpha}$ where $r_{i\alpha}$ is the reaction rate of component $i$ in phase $\alpha$. Mass is conserved in a given phase so that $\sum_{i=1}^{n_c} r_{i\alpha} = 0$, where the $r_{i\alpha}$ is in mass units. The second source term $r_{mi\alpha}$ in Eq. (1) expresses the rate of mass transfer of component $i$ from or into the phase $\alpha$ owing to vaporization/condensation and sorption. The last source term in Eq. (1) $q_{i\alpha}$ represents physical sources (wells).

The mass fraction of component $i$ in phase $\alpha$ in $V_b$ is defined to be $\omega_{i\alpha}$. The parameter $\omega_{i\alpha}$ is the mass of component $i$ in phase $\alpha$ divided by mass of the phase. Hence, $\sum_{i=1}^{n_c} \omega_{i\alpha} = 1$. With that definition,

$$\rho_{i\alpha} = \rho_\alpha \omega_{i\alpha}, \qquad i = 1, \ldots, n_c;\ \alpha = 1, \ldots, n_p + 1, \qquad (2)$$

where $\rho_\alpha$ is intrinsic mass density of phase $\alpha$.

Substitution of Eq. (2) into Eq. (1) gives:

$$\frac{\partial}{\partial t}(\phi_\alpha \rho_\alpha \omega_{i\alpha}) + \nabla \cdot (\phi_\alpha \rho_\alpha \omega_{i\alpha} \vec{u}_{i\alpha}) = \phi_\alpha r_{i\alpha} + r_{mi\alpha} + q_{i\alpha}, \qquad i = 1, \ldots, n_c;\ \alpha = 1, \ldots, n_p + 1. \qquad (3)$$

From their definitions, the volume fractions must obey the constraint $\sum_{\alpha=1}^{n_p+1} \phi_\alpha = 1$. It is well known that the porosity $\phi$ is defined as the fraction of the bulk permeable medium that is pore space, that is, the pore volume $V_p$ divided by the bulk volume $V_b$. The fact that the all fluid phases jointly fill the voids (pores) implies the relation $\sum_{\alpha=1}^{n_p} \phi_\alpha = \phi$.

The phase saturation $S_\alpha$ is defined as the fraction of the pore volume occupied by phase $\alpha$, that is, volume of phase $\alpha$ $V_\alpha$ divided by the pore volume $V_p$. The saturation of fluid phase $\alpha$ can also be defined as $S_\alpha = \phi_\alpha / \phi$. For fluid phases such as liquids and vapors, $\phi_\alpha = \phi S_\alpha$, $\alpha = 1, \ldots, n_p$, where $\phi S_\alpha$ also called the fluid content. For the solid ($s$) phase $\phi_s = 1 - \phi$, which is the grain volume divided by the bulk volume $V_b$. Eq. (3) can be split into the following two equations by noting that the porosity is $\phi = 1 - \phi_s$ and defining the fluid saturations $S_\alpha = \phi_\alpha/\phi$, $\alpha = 1, \ldots, n_p$:

$$\frac{\partial}{\partial t}(\phi S_\alpha \rho_\alpha \omega_{i\alpha}) + \nabla \cdot (\phi S_\alpha \rho_\alpha \omega_{i\alpha} \vec{u}_{i\alpha}) = \phi S_\alpha r_{i\alpha} + r_{mi\alpha} + q_{i\alpha}, \qquad i = 1, \ldots, n_c; \; \alpha = 1, \ldots, n_p, \qquad (4)$$

for the fluids, and if we fix a coordinate system in which $\vec{u}_{is} = 0$, and note that $q_{is} = 0$, then the constituent mass balance for the solid phase reduces to

$$\frac{\partial}{\partial t}\big((1-\phi)\rho_s \omega_{is}\big) = (1-\phi)r_{is} + r_{mis}, \qquad i = 1, \ldots, n_c. \qquad (5)$$

The statistical average apparent velocity of constituent $(i, \alpha)$ owing to both convection and dispersion is the sum of the barycentric velocity of phase $\alpha$ and the diffusion velocity of species $i$ in phase $\alpha$: $\vec{u}_{i\alpha} = \vec{u}_\alpha + \vec{u}'_{i\alpha}$. Since phase velocities are typically more accessible to measurement than species velocities, it is convenient to rewrite the constituent mass balance equations, Eq. (4), as

$$\frac{\partial}{\partial t}(\phi S_\alpha \rho_\alpha \omega_{i\alpha}) + \nabla \cdot (\phi S_\alpha \rho_\alpha \omega_{i\alpha} \vec{u}_\alpha) + \nabla \cdot \vec{j}_{Di\alpha} = \phi S_\alpha r_{i\alpha} + r_{mi\alpha} + q_{i\alpha}, \qquad i = 1, \ldots, n_c; \; \alpha = 1, \ldots, n_p \qquad (6)$$

where $\vec{j}_{Di\alpha} = \phi S_\alpha \rho_\alpha \omega_{i\alpha} \vec{u}'_{i\alpha}$ stands for the diffusive flux of constituent $(i, \alpha)$.

So far, the mathematical formulation of the mass conservation equations developed above is essentially the same as the standard formulation described in [36, 37]; where it differs is in the treatment of average velocity in the governing equations. Here we start to deviate from the standard formulation to prepare the equations for a sequential solution approach. The fluxes of component $i$ in phase $\alpha$ with respect to volume-averaged velocity $\vec{j}_{Di\alpha} = -\phi S_\alpha \overline{\overline{K}}_{i\alpha} \cdot \nabla(\rho_\alpha \omega_{i\alpha})$ and mass-averaged velocity $\vec{j}_{Di\alpha} = -\phi S_\alpha \rho_\alpha \overline{\overline{K}}_{i\alpha} \cdot \nabla(\omega_{i\alpha})$ owing to hydrodynamic dispersion alone were presented in [36]. The flux with respect to bulk volume-averaged velocity is proposed in this work:

$$\vec{j}_{Di\alpha} = \phi S_\alpha \rho_\alpha \omega_{i\alpha} \vec{u}'_{i\alpha} = -\overline{\overline{K}}_{i\alpha} \cdot \nabla(\phi S_\alpha \rho_\alpha \omega_{i\alpha}), \qquad i = 1, \ldots, n_c; \; \alpha = 1, \ldots, n_p. \qquad (7)$$

Two components of dispersion tensor $\overline{\overline{K}}_{i\alpha}$ for a homogeneous, isotropic permeable medium [38] are

$$(K_{xx})_{i\alpha} = \frac{D_{i\alpha}}{\tau} + \frac{\alpha_{l\alpha} u_{x\alpha}^2 + \alpha_{t\alpha}(u_{y\alpha}^2 + u_{z\alpha}^2)}{|\vec{u}_\alpha|},$$

$$(K_{xy})_{i\alpha} = \frac{(\alpha_{l\alpha} - \alpha_{t\alpha}) u_{x\alpha} u_{y\alpha}}{|\vec{u}_\alpha|}, \qquad i = 1, \ldots, n_c; \; \alpha = 1, \ldots, n_p, \qquad (8)$$

where the subscript $l$ refers to the spatial coordinate in the direction parallel, or longitudinal, to bulk flow, and $t$ is any direction perpendicular, or transverse, to $l$. $D_{i\alpha}$ is the effective binary diffusion coefficient of component $i$ in phase $\alpha$ [39], $\alpha_{l\alpha}$ and $\alpha_{t\alpha}$ are the longitudinal and transverse dispersivities, and $\tau$ is the permeable medium tortuosity.

A general set of partial differential equations (9) for the conservation of component $i$ in fluid phase $\alpha$ is obtained upon substitution of the definition for flux (Eq. (7)) into Eq. (6):

$$\frac{\partial}{\partial t}(\phi S_\alpha \rho_\alpha \omega_{i\alpha}) + \nabla \cdot (\phi S_\alpha \rho_\alpha \omega_{i\alpha} \vec{u}_\alpha) - \nabla \cdot [\bar{\bar{K}}_{i\alpha} \cdot \nabla(\phi S_\alpha \rho_\alpha \omega_{i\alpha})] \qquad \begin{matrix} i = 1, \dots, n_c; \\ \alpha = 1, \dots, n_p. \end{matrix} \qquad (9)$$
$$= \phi S_\alpha r_{i\alpha} + r_{mi\alpha} + q_{i\alpha}.$$

The term $r_{mi\alpha}$ is difficult to calculate without detailed analysis of the transport occurring within the phases. One typically simplifies the equations by using overall compositional balance equations. Overall compositional balance equations can be obtained by summing Eqs. (9) over the $n_p$ fluid phases including the constituent mass balance equations for the solid phase, Eq. (5):

$$\frac{\partial}{\partial t}\left[\phi \sum_{\alpha=1}^{n_p} S_\alpha \rho_\alpha \omega_{i\alpha} + (1-\phi)\rho_s \omega_{is}\right] + \nabla \cdot \left(\phi \sum_{\alpha=1}^{n_p} S_\alpha \rho_\alpha \omega_{i\alpha} \vec{u}_\alpha\right) -$$
$$\nabla \cdot \sum_{\alpha=1}^{n_p} [\bar{\bar{K}}_{i\alpha} \cdot \nabla(\phi S_\alpha \rho_\alpha \omega_{i\alpha})] = R_i, \qquad i = 1, \dots, n_c, \qquad (10)$$

where the source terms $R_i$ are a combination of all rate terms for a particular component and may be expressed as

$$R_i = \phi \sum_{\alpha=1}^{n_p} S_\alpha r_{i\alpha} + (1-\phi)r_{is} + Q_i, \qquad i = 1, \dots, n_c, \qquad (11)$$

and $Q_i = \sum_{\alpha=1}^{n_p} q_{i\alpha}$ is the injection/production rate for component $i$ per bulk volume. We have $\sum_{\alpha=1}^{n_p+1} r_{mi\alpha} = 0$, a relation following from the inability to accumulate mass at a volumeless phase interface.

There is no doubt that the Eq. (10) is true. In the currently used chemical compositional model, the final form of Eq. (10) is expressed in terms of volume fraction of component $i$ in phase $\alpha$, $c_{i\alpha}$, under the assumption of ideal mixing. To account for the reduction in pore volume caused by adsorption, the coefficient $\left(1 - \sum_{i=1}^{n_{cv}} \hat{c}_i\right)$ is introduced into the overall compositional balance equations in the currently used chemical compositional model [34 and 40], where $\hat{c}_i$ is the adsorbed concentration of species $i$, and $n_{cv}$ is the total number of volume-occupying components. The coefficient represents reduction in pore volume due to adsorption. During the process of this research, it was revealed that even though this approach estimates the adsorption effect on the transport of a component reasonably well, it does not satisfy the species-conservation equation since the coefficient is introduced only in the accumulation term of the overall compositional balance equations. It is well known that an equation remains balanced when both sides of an equation are multiplied by the same nonzero quantity. This mathematical inequality is a clear indication of that the law of conservation of mass is violated. The energy conservation equation does not consider any change in the porosity caused by adsorption at all. This approach leads to inconsistency between the overall compositional balance equations and the energy conservation equation in the currently used chemical compositional model by violating conservation of energy.

We state that not only accumulation term in the mass-balance equation should be modified due to adsorption effect, but also other terms in the model that include the volume fraction of a phase. In order to be more convincing, let us start with the derivation of a simplified mass-balance model which accounts for the reduction in the porosity and the change in volume fractions of the fluid and solid phases caused by adsorption.

Consider the interphase mass transfer process in a bulk volume $V_b$ at the representative elementary volume scale with no convection-diffusion-dispersion-reaction and no injection/production. Under these assumptions, Eq. (1) reduces to

$$\frac{d}{dt}(\phi_\alpha \rho_{i\alpha}) = r_{mi\alpha}, \qquad i = 1, \ldots, n_c;\ \alpha = 1, \ldots, n_p + 1. \qquad (12)$$

In general, the source term $r_{mi\alpha}$ in Eq. (12) expresses the rate of mass transfer of component $i$ from or into the phase $\alpha$ owing to vaporization, condensation, and sorption.

Equation (12) can be separated into two equations as

$$\frac{d}{dt}(\phi_\alpha \rho_{i\alpha}) = r_{mi\alpha}, \qquad i = 1, \ldots, n_c;\ \alpha = 1, \ldots, n_p \qquad (13)$$

for the fluid phases, and

$$\frac{d}{dt}(\phi_s \rho_{is}) = r_{mis}, \qquad i = 1, \ldots, n_c, \qquad (14)$$

for the solid phase.

Summing the equation (13) over all fluid phases $\alpha$ and using the constraint $\sum_{\alpha=1}^{n_p} r_{mi\alpha} + r_{mis} = 0$, we obtain the following system of equations:

$$\frac{d}{dt}\left[\sum_{\alpha=1}^{n_p}(\phi_\alpha \rho_{i\alpha})\right] = -r_{mis}, \qquad i = 1, \ldots, n_c, \qquad (15)$$

$$\frac{d}{dt}(\phi_s \rho_{is}) = r_{mis}, \qquad i = 1, \ldots, n_c. \qquad (16)$$

The source terms those remain in the system of equations (15) and (16) allow for mass transfer by adsorption. They are equal in absolute value but opposite in sign. In other words, the net loss of a component by fluid phases is the gain of the same component by solid phase. A good way to understand this is to consider a microscopic view of pores and grains within a medium. Figure 1 illustrates a cube of small volume placed within a permeable medium.

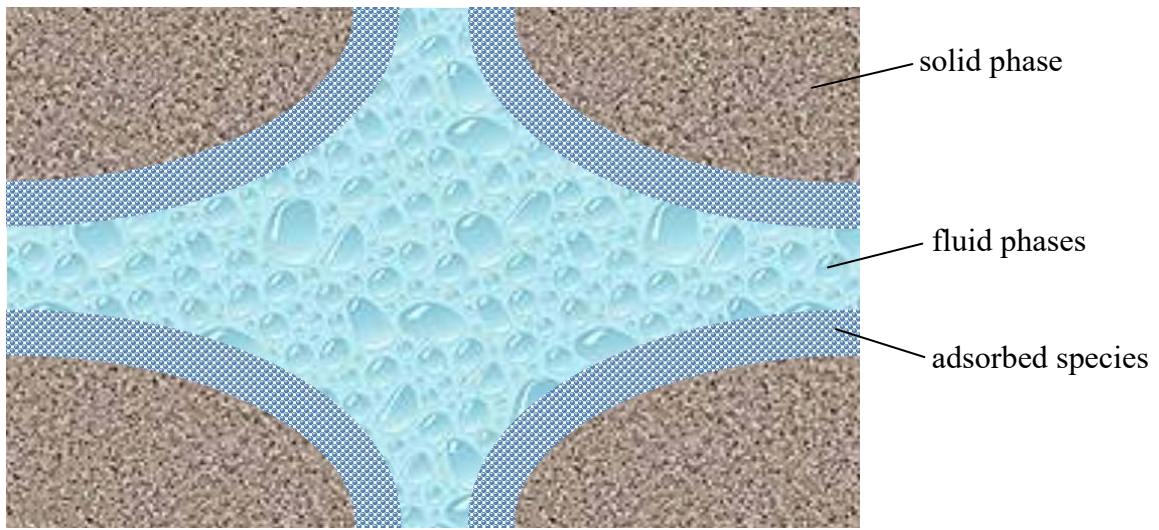

**Figure 1** Illustration of microscopic cube placed within a permeable medium

Note that the equation (15) obtained as a result of summing over all fluid phases, unlike the equation (13), does not describe the mass transfer processes taking place between the fluid phases. Because when we sum the mass transfer rates over all fluid phases, gains and losses owing to mass transfer between fluid phases cancel each other out. Flash calculations are used to describe mass transfer between the fluid phases such as phase condensation and evaporation due to changes in temperature and pressure.

Adding the two equations in the last system (Eqs. (15) and (16)) together yields

$$\frac{d}{dt}\left[\sum_{\alpha=1}^{n_p}(\phi_\alpha \rho_{i\alpha}) + \phi_s \rho_{is}\right] = 0, \qquad i = 1, \dots, n_c. \tag{17}$$

Since the time derivative of the sum is zero, the sum remains constant over time. That is, the total mass of the isolated system is independent of any changes taking place within the system during the process of adsorption, which corresponds to the conservation of mass in classical continuum mechanics. Note that this does not imply that the volume fraction of each phase and intrinsic mass density of each constituent are constants. The $2n_p + 2$ quantities in the equation (17) vary in such a manner that the sum remains constant in time for an isolated system. Applying the equation (17) at any particular instant of time, we can rewrite it as follows:

$$\frac{d}{dt}\left[\sum_{\alpha=1}^{n_p}(\breve{\phi}_\alpha \breve{\rho}_{i\alpha}) + \hat{\phi}_s \hat{\rho}_{is}\right] = 0, \qquad i = 1, \dots, n_c, \tag{18}$$

where the breve ($\breve{\ }$) and hat ($\hat{\ }$) notations denote the adsorption affected physical properties of the fluid phases and the solid phase, respectively.

The adsorption effect can be introduced into the overall compositional balance equation with convection, reaction and physical source terms in a similar way as in the derivation above. The overall compositional balance equation is derived from conservation principles of mass via an integral solution relation stating that the sum of the changes of component's mass within a control volume $V$ (Fig. 2) must be equal to the component's resultant mass is lost (or gained) through the bounding surface area $A$ of the volume plus total mass of the component generated/consumed by sources and sinks inside the control volume.

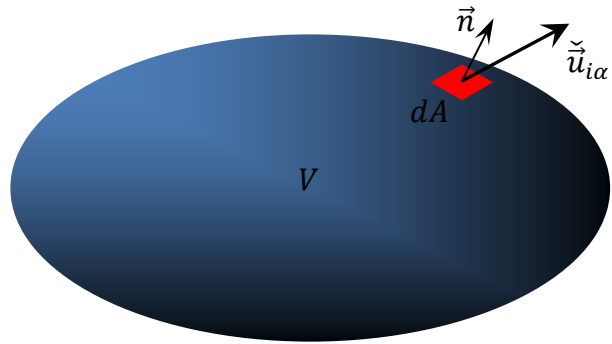

**Figure 2** The closed surface $A$ defines the volume $V$. The unit vector $\vec{n}$ is the outward normal to the surface element $dA$, and $\vec{\breve{u}}_{i\alpha}$ is the average pore velocity vector of constituent $(i, \alpha)$ at the surface element affected by adsorption.

Since in adsorption-related processes local transport and reaction take place in the pore space changed due to adsorption, this is expressed by the following integral continuity equation:

$$\frac{d}{dt}\left\{\int_V \left[\sum_{\alpha=1}^{n_p}(\check{\phi}_\alpha\check{\rho}_{i\alpha}) + \hat{\phi}_s\hat{\rho}_{is}\right]dV\right\} + \oint_A \sum_{\alpha=1}^{n_p}(\check{\phi}_\alpha\check{\rho}_{i\alpha}\check{\vec{u}}_{i\alpha}) \cdot \vec{n}\, dA$$

$$= \int_V \left[\sum_{\alpha=1}^{n_p}(\check{\phi}_\alpha\check{r}_{i\alpha}) + \hat{\phi}_s\hat{r}_{is}\right]dV + \int_V Q_i dV, \qquad i = 1, \ldots, n_c. \qquad (19)$$

where $\check{r}_{i\alpha}$ and $\hat{r}_{is}$ denote the adsorption affected reaction rates of component $i$ in the fluid phases and in the solid phase, respectively.

The divergence theorem may be applied to the surface integral, changing it into a volume integral:

$$\oint_A \sum_{\alpha=1}^{n_p}(\check{\phi}_\alpha\check{\rho}_{i\alpha}\check{\vec{u}}_{i\alpha}) \cdot \vec{n}\, dA = \int_V \nabla \cdot \sum_{\alpha=1}^{n_p}(\check{\phi}_\alpha\check{\rho}_{i\alpha}\check{\vec{u}}_{i\alpha})\, dV, \qquad i = 1, \ldots, n_c. \qquad (20)$$

Applying Leibniz's rule to the integral on the left

$$\frac{d}{dt}\left\{\int_V \left[\sum_{\alpha=1}^{n_p}(\check{\phi}_\alpha\check{\rho}_{i\alpha}) + \hat{\phi}_s\hat{\rho}_{is}\right]dV\right\} = \int_V \frac{\partial}{\partial t}\left[\sum_{\alpha=1}^{n_p}(\check{\phi}_\alpha\check{\rho}_{i\alpha}) + \hat{\phi}_s\hat{\rho}_{is}\right]dV, \qquad i = 1, \ldots, n_c, \qquad (21)$$

and then combining all of the integrals:

$$\int_V \left\{\frac{\partial}{\partial t}\left[\sum_{\alpha=1}^{n_p}(\check{\phi}_\alpha\check{\rho}_{i\alpha}) + \hat{\phi}_s\hat{\rho}_{is}\right] + \nabla \cdot \sum_{\alpha=1}^{n_p}(\check{\phi}_\alpha\check{\rho}_{i\alpha}\check{\vec{u}}_{i\alpha}) - \left[\sum_{\alpha=1}^{n_p}(\check{\phi}_\alpha\check{r}_{i\alpha}) + \hat{\phi}_s\hat{r}_{is}\right] - Q_i\right\}dV = 0, \qquad i = 1, \ldots, n_c. \qquad (22)$$

The integral must be zero for any control volume; this can only be true if the integrand itself is zero, so that:

$$\frac{\partial}{\partial t}\left[\sum_{\alpha=1}^{n_p}(\check{\phi}_\alpha\check{\rho}_{i\alpha}) + \hat{\phi}_s\hat{\rho}_{is}\right] + \nabla \cdot \sum_{\alpha=1}^{n_p}(\check{\phi}_\alpha\check{\rho}_{i\alpha}\check{\vec{u}}_{i\alpha})$$

$$= \sum_{\alpha=1}^{n_p}(\check{\phi}_\alpha\check{r}_{i\alpha}) + \hat{\phi}_s\hat{r}_{is} + Q_i, \qquad i = 1, \ldots, n_c. \qquad (23)$$

Recall that by definition

$$\sum_{\alpha=1}^{n_p}\phi_\alpha + \phi_s = 1, \quad \phi = \sum_{\alpha=1}^{n_p}\phi_\alpha, \quad \phi_s = 1 - \phi, \qquad (24)$$

and adsorption is described through isotherms. It is assumed that the all fluid phases and the solid phase jointly fill the bulk volume, regardless of whether an adsorption layer exists or not. This implies the relation

$$\sum_{\alpha=1}^{n_p} \check{\phi}_\alpha + \hat{\phi}_s = 1. \tag{25}$$

Now, similarly to the classical case, we define

$$\hat{\phi} = \sum_{\alpha=1}^{n_p} \check{\phi}_\alpha, \quad \check{S}_\alpha = \frac{\check{\phi}_\alpha}{\hat{\phi}}, \qquad \alpha = 1, \ldots, n_p, \tag{26}$$

and

$$\hat{\phi}_s = 1 - \hat{\phi}, \tag{27}$$

where the modified porosity $\hat{\phi}$ is defined as the fraction of the bulk permeable medium that is pore space remaining after adsorption, and the modified phase saturation $\check{S}_\alpha$ is defined as the fraction of the reduced pore volume occupied by phase $\alpha$.

The average pore velocity vector of constituent $(i, \alpha)$ owing to both convection and dispersion is

$$\check{\vec{u}}_{i\alpha} = \check{\vec{u}}_\alpha - \frac{\bar{\bar{\check{K}}}_{i\alpha} \cdot \nabla(\check{\phi}_\alpha \check{\rho}_{i\alpha})}{\check{\phi}_\alpha \check{\rho}_{i\alpha}}, \qquad \alpha = 1, \ldots, n_p. \tag{28}$$

The definitions for the volume fractions and the average pore velocity vector of constituent $(i, \alpha)$ above substituted into Eq. (23) give

$$\frac{\partial}{\partial t}\left[\hat{\phi} \sum_{\alpha=1}^{n_p}(\check{S}_\alpha \check{\rho}_{i\alpha}) + \hat{\phi}_s \hat{\rho}_{is}\right] + \nabla \cdot \left(\hat{\phi} \sum_{\alpha=1}^{n_p} \check{S}_\alpha \check{\rho}_{i\alpha} \check{\vec{u}}_\alpha\right) - $$
$$\nabla \cdot \sum_{\alpha=1}^{n_p}\left[\bar{\bar{\check{K}}}_{i\alpha} \cdot \nabla(\hat{\phi} \check{S}_\alpha \check{\rho}_{i\alpha})\right] = \hat{\phi} \sum_{\alpha=1}^{n_p} \check{S}_\alpha \check{r}_{i\alpha} + \hat{\phi}_s \hat{r}_{is} + Q_i, \qquad i = 1, \ldots, n_c. \tag{29}$$

The dispersion tensor $\bar{\bar{\check{K}}}_{i\alpha}$ is treated in a similar manner to that described above. Note that equation (29) is just an expansion of equation (10) in terms of porosity changes due to adsorption. Since $\hat{\phi} + \hat{\phi}_s = 1$, let us assume that the change in the porosity and volume fraction of the solid phase due to the adsorption can be represented as

$$\hat{\phi} = \phi - \phi' \tag{30}$$

and

$$\hat{\phi}_s = \phi_s + \phi', \tag{31}$$

respectively, where $\phi'$ denotes the volume fraction of the components adsorbed onto the solid surface.

In order to keep the equations of our proposed model formulation as similar to the equations used in the currently used chemical compositional model formulation as possible, we model the volume fraction of the components adsorbed onto the solid surface as

$$\phi' = \phi \sum_{i=1}^{n_{cv}} \hat{c}_i, \tag{32}$$

where $\hat{c}_i$ is the adsorbed component concentration, measured as volume of component $i$ in phase $\alpha$ per unit pore volume. An adsorption isotherm is used to calculate the adsorbed concentration of component $i$, $\hat{c}_i$. Hence, the modified porosity $\hat{\phi}$ is related to the original porosity $\phi$ as follows:

$$\hat{\phi} = \phi \left(1 - \sum_{i=1}^{n_{cv}} \hat{c}_i \right). \tag{33}$$

Let $\check{\rho}_{ii\alpha}$ and $\hat{\rho}_{iis}$ denote the densities of component $i$ in the fluid phases and in the solid phase, respectively. The parameter $\check{\rho}_{ii\alpha}$ is the mass of component $i$ in phase $\alpha$ divided by the volume of component $i$ in the same phase. By analogy, $\hat{\rho}_{iis}$ is defined as the mass of component $i$ in the solid phase divided by the volume of component $i$ in the solid phase. We now define the modified volume fractions of component $i$ in the fluid phases and in the solid phase due to adsorption to be

$$\check{c}_{i\alpha} = \check{\rho}_{i\alpha}/\check{\rho}_{ii\alpha}, \qquad i = 1, \ldots, n_c; \alpha = 1, \ldots, n_p, \tag{34}$$

and

$$\hat{c}_{is} = \hat{\rho}_{is}/\hat{\rho}_{iis}, \qquad i = 1, \ldots, n_c, \tag{35}$$

respectively. The content of component $i$ in the solid phase can be expressed in terms of volume of component $i$ adsorbed per unit pore volume as

$$\hat{c}_{is} = \phi \hat{c}_i / \hat{\phi}_s, \qquad i = 1, \ldots, n_c. \tag{36}$$

Assuming that density of component $i$ is the same in all phases, we have:

$$\check{\rho}_{ii\alpha} = \hat{\rho}_{iis} = \rho_i, \qquad i = 1, \ldots, n_c; \alpha = 1, \ldots, n_p, \tag{37}$$

where $\rho_i$ is the pure component mass density in units of mass of component $i$ per unit volume of component $i$.

Using Eqs. (30)-(37), we can write Eq. (29) as

$$\frac{\partial}{\partial t}\left\{\phi \rho_i \left[\left(1 - \sum_{j=1}^{n_{cv}} \hat{c}_j\right) \sum_{\alpha=1}^{n_p} \check{S}_\alpha \check{c}_{i\alpha} + \hat{c}_i\right]\right\} + \nabla \cdot \left(\hat{\phi} \rho_i \sum_{\alpha=1}^{n_p} \check{S}_\alpha \check{c}_{i\alpha} \vec{\check{u}}_\alpha\right) - \nabla \cdot \left\{\sum_{\alpha=1}^{n_p} \left[\overline{\overline{\check{K}}}_{i\alpha} \cdot \nabla(\hat{\phi} \rho_i \check{S}_\alpha \check{c}_{i\alpha})\right]\right\} = R_i, \qquad i = 1, \ldots, n_c. \tag{38}$$

where

$$R_i = \hat{\phi} \sum_{\alpha=1}^{n_p} \check{S}_\alpha \check{r}_{i\alpha} + (1-\hat{\phi})\hat{r}_{is} + Q_i, \qquad i = 1, \ldots, n_c. \qquad (39)$$

There are no general functions for $\check{r}_{i\alpha}$ and $\hat{r}_{is}$. An example of a first-order reaction rate for radioactive decay or biodegradation is

$$\check{r}_{i\alpha} = -k_i \check{\rho}_{i\alpha}, \qquad i = 1, \ldots, n_c; \; \alpha = 1, \ldots, n_p + 1, \qquad (40)$$

and

$$\hat{r}_{is} = -k_i \hat{\rho}_{is}, \qquad i = 1, \ldots, n_c, \qquad (41)$$

where $k_i$ is the decay constant or reaction rate coefficient in units of inverse time. With these, the source term, Eq. (39), become

$$R_i = -k_i \phi \rho_i \left[ \left(1 - \sum_{j=1}^{n_{cv}} \hat{c}_j\right) \sum_{\alpha=1}^{n_p} \check{S}_\alpha \check{c}_{i\alpha} + \hat{c}_i \right] + Q_i, \qquad i = 1, \ldots, n_c. \qquad (42)$$

In order to keep similar notations as in the currently used chemical compositional model formulation for easier comparison of the formulations, further we omit the notations denoting the adsorption affected physical properties except the hat (^) notation for the adsorption affected solid phase properties $\hat{\phi}$ and $\hat{c}_i$.

We define the overall concentration $\tilde{c}_i$ as

$$\tilde{c}_i = \left(1 - \sum_{j=1}^{n_{cv}} \hat{c}_j\right) \sum_{\alpha=1}^{n_p} S_\alpha c_{i\alpha} + \hat{c}_i, \qquad i = 1, \ldots, n_c. \qquad (43)$$

The fact that all species, including both the components in the flowing phases, and the components in the stationary phase (adsorbed on the rock surface), jointly fill the pore volume implies the relation

$$\sum_{i=1}^{n_c} \tilde{c}_i = 1. \qquad (44)$$

Now Eq. (38) can be rewritten as

$$\frac{\partial}{\partial t}(\phi \rho_i \tilde{c}_i) + \nabla \cdot \left[\hat{\phi} \rho_i \sum_{\alpha=1}^{n_p}(S_\alpha c_{i\alpha} \vec{u}_\alpha)\right] - \nabla \cdot \sum_{\alpha=1}^{n_p} [\bar{\bar{K}}_{i\alpha} \cdot \nabla(\hat{\phi} \rho_i S_\alpha c_{i\alpha})] = R_i, \qquad i = 1, \ldots, n_c, \qquad (45)$$

where

$$R_i = \hat{\phi} \sum_{\alpha=1}^{n_p} S_\alpha r_{i\alpha} + (1-\hat{\phi}) r_{is} + Q_i, \qquad i = 1, \ldots, n_c, \qquad (46)$$

or for biodegradation

$$R_i = -k_i \phi \rho_i \left[ \left(1 - \sum_{j=1}^{n_{cv}} \hat{c}_j \right) \sum_{\alpha=1}^{n_p} S_\alpha c_{i\alpha} + \hat{c}_i \right] + Q_i, \qquad i = 1, \ldots, n_c. \qquad (47)$$

The derivation of this mass-balance model which accounts for the change in the porosity and volume fraction of the solid phase caused by adsorption proves that the species-conservation equations used in the currently used chemical compositional model [33, 34, 40, and 41] violate the principle of mass conservation. One can be convinced of the validity of this conclusion by comparing the reaction rate terms of the species conservation equations in our formulation (Eq. (46)) with those of the currently used chemical compositional model formulation. Since in our approach the factor $\left(1 - \sum_{i=1}^{n_{cv}} \hat{c}_i\right)$ representing the change in pore volume caused by adsorption is introduced as a multiplier in the modified porosity calculation formula (Eq. (33)), according to the multiplication property of equality, the equality in Eq. (45) is preserved. Note that the new and existing formulations agree with each other when $\sum_{i=1}^{n_{cv}} \hat{c}_i \to 0$.

Variation of pore volume with pore pressure can be taken into account by the pressure dependence of porosity. The porosity depends on pressure due to rock compressibility, which is often assumed to be constant. Therefore,

$$\phi = \phi_R [1 + c_r (p_1 - p_S)], \qquad (48)$$

where $\phi_R$ is the porosity at a specific pressure $p_S$, $p_1$ is the water phase pressure, and $c_r$ is the rock compressibility at $p_S$.

A slightly compressible fluid has a small but constant compressibility. For a slightly compressible fluid, the component density $\rho_i$ can be written as:

$$\rho_i = \rho_{iR} [1 + c_i^0 (p_1 - p_R)], \qquad i = 1, \ldots, n_c, \qquad (49)$$

where $\rho_{iR}$ is the density of component $i$ at the standard pressure $p_R$, a constant value. $c_i^0$ is the compressibility of component $i$.

Since reference density $\rho_{iR}$ is constant for each component we can divide through both sides of Eq. (45) by $\rho_{iR}$. In terms of the dimensionless density $\bar{\rho}_i = \rho_i / \rho_{iR}$ Eq. (45) can be written as:

$$\frac{\partial (\phi \bar{\rho}_i \tilde{c}_i)}{\partial t} + \nabla \cdot \hat{\phi} \bar{\rho}_i \sum_{\alpha=1}^{n_p} (S_\alpha c_{i\alpha} \vec{u}_\alpha) - \nabla \cdot \sum_{\alpha=1}^{n_p} \left[ \bar{\bar{K}}_{i\alpha} \cdot \nabla (\hat{\phi} \bar{\rho}_i S_\alpha c_{i\alpha}) \right] \\ = -k_i \phi \bar{\rho}_i \tilde{c}_i + \frac{Q_i}{\rho_{iR}}, \qquad i = 1, \ldots, n_c. \qquad (50)$$

The phase velocity from Darcy's law is

$$\vec{u}_\alpha = -\frac{\bar{\bar{k}} k_{r\alpha}}{\hat{\phi} S_\alpha \mu_\alpha} (\nabla p_\alpha - \gamma_\alpha \nabla z), \qquad \alpha = 1, \ldots, n_p, \qquad (51)$$

where $\bar{\bar{k}}$ is the permeability tensor, $k_{r\alpha}$ is the relative permeability of fluid phase $\alpha$, $\mu_\alpha$ is the dynamic viscosity of fluid phase $\alpha$, $p_\alpha$ is the pressure in fluid phase $\alpha$, $\gamma_\alpha$ is the specific weight for fluid phase $\alpha$, and $z$ represents depth.

We sum the overall compositional balance equations (Eq. (50)) over the $n_c$ components to obtain the equation of continuity, or conservation of total mass. The equation of continuity is

$$\phi_R c_t \frac{\partial p_1}{\partial t} + \Delta_t F(\tilde{c}_i) + \nabla \cdot \left\{ \hat{\phi} \sum_{\alpha=1}^{n_p} \left( S_\alpha \vec{u}_\alpha \sum_{i=1}^{n_c} \bar{\rho}_i c_{i\alpha} \right) \right\} = \sum_{i=1}^{n_c} \frac{Q_i}{\rho_{iR}}, \tag{52}$$

where we used that

$$\sum_{i=1}^{n_c} \nabla \cdot \vec{j}_{Di\alpha} = 0, \qquad \alpha = 1, \dots, n_p, \tag{53}$$

(net dispersive flux in a phase is zero), or

$$\sum_{i=1}^{n_c} \nabla \cdot [\bar{\bar{K}}_{i\alpha} \cdot \nabla(\hat{\phi} \bar{\rho}_i S_\alpha c_{i\alpha})] = 0, \qquad \alpha = 1, \dots, n_p, \tag{54}$$

and according to the total reaction definition

$$\sum_{i=1}^{n_c} \left( \sum_{\alpha=1}^{n_p} R_{i\alpha} + R_{is} \right) = 0, \tag{55}$$

we have

$$\phi \sum_{i=1}^{n_c} (k_i \bar{\rho}_i \tilde{c}_i) = 0. \tag{56}$$

The total compressibility, $c_t$, is

$$c_t = c_r + [1 + c_r(2p_1 - p_S - p_R)] \sum_{i=1}^{n_c} (c_i^0 \tilde{c}_i), \tag{57}$$

and the term

$$\Delta_t F(\tilde{c}_i) = \phi(p_1 - p_R) \sum_{i=1}^{n_c} \left( c_i^0 \frac{\partial \tilde{c}_i}{\partial t} \right) \tag{58}$$

can be treated as a known (source type) function that is determined using the values from the previous time step.

The pressure equation is developed by substituting Darcy's law (Eq. (51)) for the phase flux term of Eq. (52), using the definition of capillary pressure $p_{c\alpha 1} = p_\alpha - p_1, \alpha = 2, \dots, n_p$. The pressure equation in terms of the reference phase (phase 1) pressure is

$$\phi_R c_t \frac{\partial p_1}{\partial t} - \nabla \cdot (\bar{\bar{k}} \lambda_{rTc} \nabla p_1)$$
$$= \nabla \cdot \left( \bar{\bar{k}} \sum_{\alpha=1}^{n_p} \lambda_{r\alpha c} \nabla p_{c\alpha 1} \right) - \nabla \cdot \left( \bar{\bar{k}} \sum_{\alpha=1}^{n_p} (\lambda_{r\alpha c} \gamma_\alpha) \nabla z \right) - \Delta_t F(\tilde{c}_i) + \sum_{i=1}^{n_c} \frac{Q_i}{\rho_{iR}}, \quad (59)$$

where

$$\lambda_{r\alpha c} = \lambda_{r\alpha} \sum_{i=1}^{n_c} \bar{\rho}_i c_{i\alpha}, \quad \lambda_{r\alpha} = \frac{k_{r\alpha}}{\mu_\alpha}, \quad \alpha = 1, \ldots, n_p, \quad (60)$$

and total relative mobility is

$$\lambda_{rTc} = \sum_{\alpha=1}^{n_p} \lambda_{r\alpha c}. \quad (61)$$

In order to implement a sequential solution approach to solve the overall compositional balance equations (Eq. (45)) implicitly, by applying the mean value estimate for character sums in Eq. (45), we can write

$$\sum_{\alpha=1}^{n_p} (S_\alpha c_{i\alpha} \vec{u}_\alpha) = \vec{u}_i \sum_{\alpha=1}^{n_p} (S_\alpha c_{i\alpha}), \quad i = 1, \ldots, n_c, \quad (62)$$

and

$$\sum_{\alpha=1}^{n_p} [\bar{\bar{K}}_{i\alpha} \cdot \nabla(\hat{\phi}\rho_i S_\alpha c_{i\alpha})] = \bar{\bar{K}}_i \cdot \sum_{\alpha=1}^{n_p} \nabla(\hat{\phi}\rho_i S_\alpha c_{i\alpha}), \quad i = 1, \ldots, n_c, \quad (63)$$

where $\vec{u}_i$ and $\bar{\bar{K}}_i$ can be defined as some averages. Since differentiation and summation are interchangeable operations in this system, the sum of the gradients can be calculated as the gradient of the sum

$$\bar{\bar{K}}_i \cdot \sum_{\alpha=1}^{n_p} \nabla(\hat{\phi}\rho_i S_\alpha c_{i\alpha}) = \bar{\bar{K}}_i \cdot \nabla\left[\hat{\phi}\rho_i \sum_{\alpha=1}^{n_p} (S_\alpha c_{i\alpha})\right], \quad i = 1, \ldots, n_c. \quad (64)$$

Equation (45) can be written using Eqs. (62), (63), and (64) as below:

$$\frac{\partial}{\partial t}\left\{\hat{\phi}\rho_i\left[\left(1 - \sum_{j=1}^{n_{cv}} \hat{c}_j\right)c_i + \hat{c}_i\right]\right\} + \nabla \cdot (\hat{\phi}\rho_i \vec{u}_i c_i) - \nabla \cdot [\bar{\bar{K}}_i \cdot \nabla(\hat{\phi}\rho_i c_i)]$$
$$= -k_i \hat{\phi}\rho_i\left[\left(1 - \sum_{j=1}^{n_{cv}} \hat{c}_j\right)c_i + \hat{c}_i\right] + Q_i, \quad i = 1, \ldots, n_c, \quad (65)$$

where the total concentration of component $i$ in fluid phases is

$$c_i = \sum_{\alpha=1}^{n_p} (S_\alpha c_{i\alpha}), \qquad i = 1, \dots, n_c. \tag{66}$$

This new mathematical formulation of species conservation equations makes it possible to apply a sequential solution approach to solve each of these equations separately and implicitly for the total concentration of each component in fluid phases, $c_i$. A flash calculation is then performed to obtain the phase saturations and the concentrations of components in each phase. The linear, Freundlich or Langmuir adsorption isotherm model can be applied to calculate the adsorbed concentrations $\hat{c}_i$. Numerical values of $\vec{\tilde{u}}_i$ and $\overline{\overline{\tilde{K}}}_i$ can be most simply calculated as the weighted averages

$$\vec{\tilde{u}}_i = \frac{\sum_{\alpha=1}^{n_p} S_\alpha c_{i\alpha} \vec{u}_\alpha}{c_i}, \qquad \overline{\overline{\tilde{K}}}_i = \frac{\sum_{\alpha=1}^{n_p} \overline{\overline{K}}_{i\alpha} \cdot \nabla(\hat{\phi} \rho_i S_\alpha c_{i\alpha})}{\nabla(\hat{\phi} \rho_i c_i)}, \qquad i = 1, \dots, n_c, \tag{67}$$

obtained from previous time-step values.
The energy balance equation is derived by assuming that energy is a function of temperature only and energy flux in the aquifer or reservoir occurs by advection and heat conduction only. We write the energy conservation equation in terms of the modified porosity $\hat{\phi}$, as opposed to the existing chemical compositional model formulation. The energy conservation equation reads

$$\frac{\partial}{\partial t} \left[ \hat{\phi} \sum_{\alpha=1}^{n_p} \rho_\alpha S_\alpha C_{V\alpha} + (1-\hat{\phi})\rho_s C_s \right] T + \nabla \cdot \left( \hat{\phi} \sum_{\alpha=1}^{n_p} \rho_\alpha S_\alpha C_{p\alpha} \vec{u}_\alpha T - k_T \nabla T \right) = q_c - q_L, \tag{68}$$

where $T$ is the temperature; $C_{V\alpha}$ and $C_{p\alpha}$ are the heat capacities of phase $\alpha$ at constant volume and pressure, respectively; $C_s$ is the heat capacity of the solid phase; $k_T$ is the thermal conductivity; $q_c$ is the heat source term; and $q_L$ is the heat loss to overburden and underburden formations or soil.
The sequential solution procedure is carried out in the following order: (a) solution of the pressure equation (59) implicitly, (b) solution of the transport system (Eq. (65)) implicitly for the total concentration of each component, and (c) solution of the energy conservation equation (68) implicitly.
Table 1 summarizes the key differences between the currently used chemical compositional model and the newly developed model formulations.

**Table 1** A comparison of the key differences between the currently used chemical compositional model and the new model formulations

| CURRENTLY USED FORMULATION | NEW FORMULATION |
|---|---|
| The fluxes of component $i$ in phase $\alpha$ are expressed with respect to mass-averaged velocity $$\vec{J}_{Di\alpha} = -\phi S_\alpha \rho_\alpha \bar{\bar{K}}_{i\alpha} \cdot \nabla(\omega_{i\alpha}).$$ | The flux with respect to bulk volume-averaged velocity is proposed in this work: $$\vec{J}_{Di\alpha} = -\bar{\bar{K}}_{i\alpha} \cdot \nabla(\phi S_\alpha \rho_\alpha \omega_{i\alpha}).$$ |
| The reaction rate term of component $i$ is written in terms of the original porosity $\phi$: $$R_i = \phi \sum_{\alpha=1}^{n_p} S_\alpha r_{i\alpha} + (1-\phi)r_{is} + Q_i.$$ | The reaction rate term of component $i$ is written in terms of the modified porosity $\hat{\phi}$: $$R_i = \hat{\phi} \sum_{\alpha=1}^{n_p} S_\alpha r_{i\alpha} + (1-\hat{\phi})r_{is} + Q_i.$$ |
| The model is based on the governing equations with the superficial Darcy velocity of the phase. | The governing equations are written in terms of volume-averaged pore velocity of the phase. The volume-averaged pore velocity is related to the superficial Darcy velocity by the fluid content $\phi S_\alpha$. |
| Energy conservation equation is written in terms of the original porosity $\phi$. | Energy conservation equation is written in terms of the modified porosity $\hat{\phi}$. |

**Results and Discussion**

The critical stage of the model development process is the evaluation of whether or not a given mathematical model describes a phenomenon accurately. At the current stage of the development process our main goal is to evaluate whether our recently developed mathematical formulation is appropriate for simulating the mechanism of ASP flooding.

Since there is no analytical solution available for the chemical compositional problem under consideration, the verification test is based on the comparison with simulator, UTCHEM, for the case when change in porosity due to the adsorption does not influence the process significantly. The comparative studies below show that the results obtained from IMPEC implementation of the newly proposed formulation are in a good agreement with that of UTCHEM simulator. We use S3GRAF software, developed and licensed by Sciencesoft Ltd., for post-processing the output data. The number of grids used is 7×18×89 (see Fig. 3).

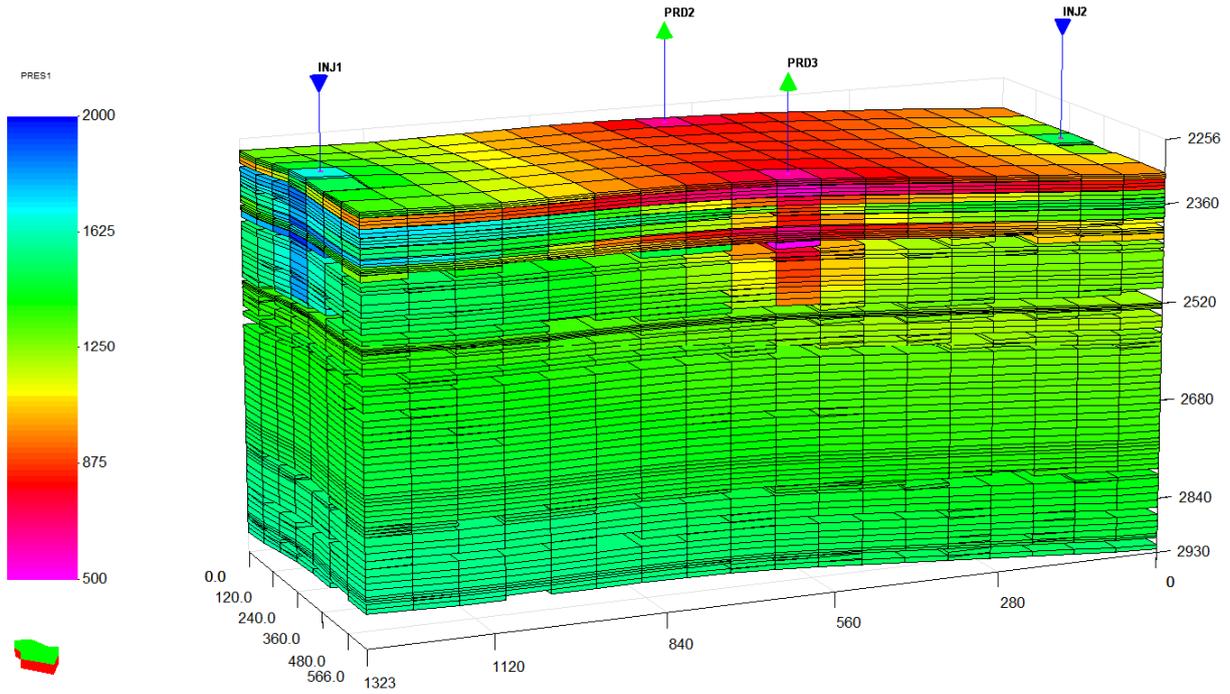

**Figure 3** The computational domain

The matches between the currently used and the newly developed chemical compositional model formulations' numerical results for the matched variables are shown in Figs. 4 and 5 for the injected pore volume in the range 0–0.012 PV.

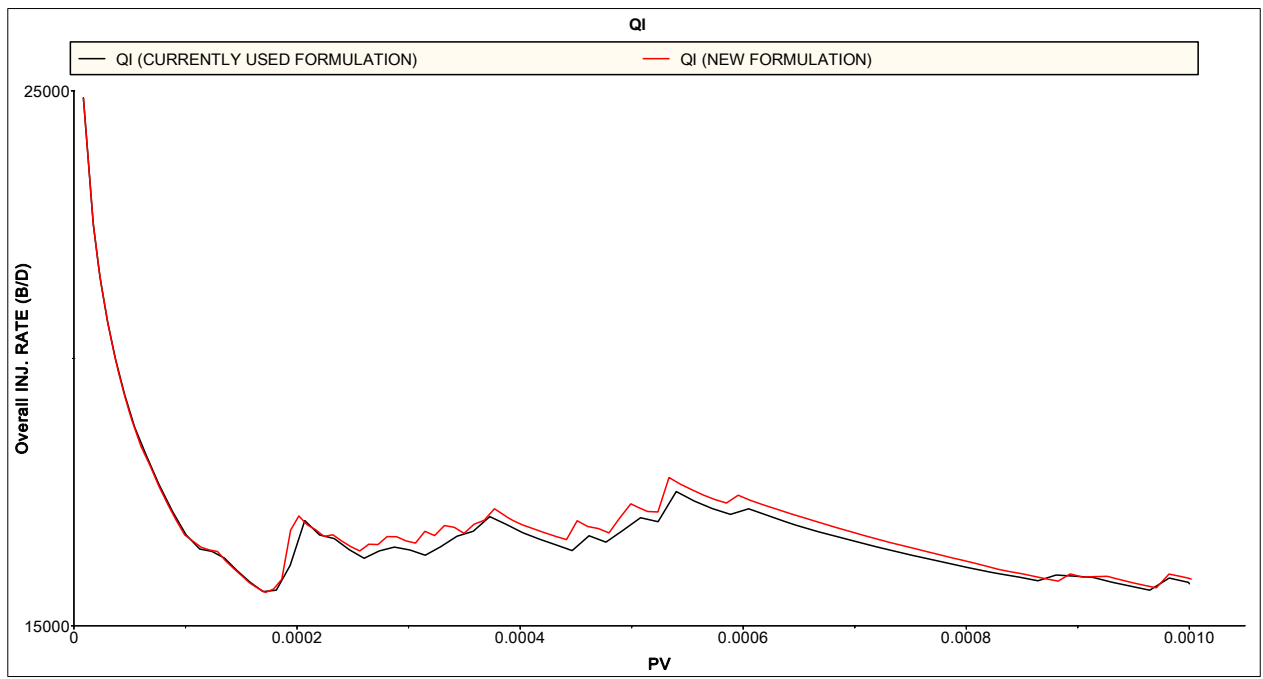

**Figure 4** Overall injection rate vs. injected pore volume

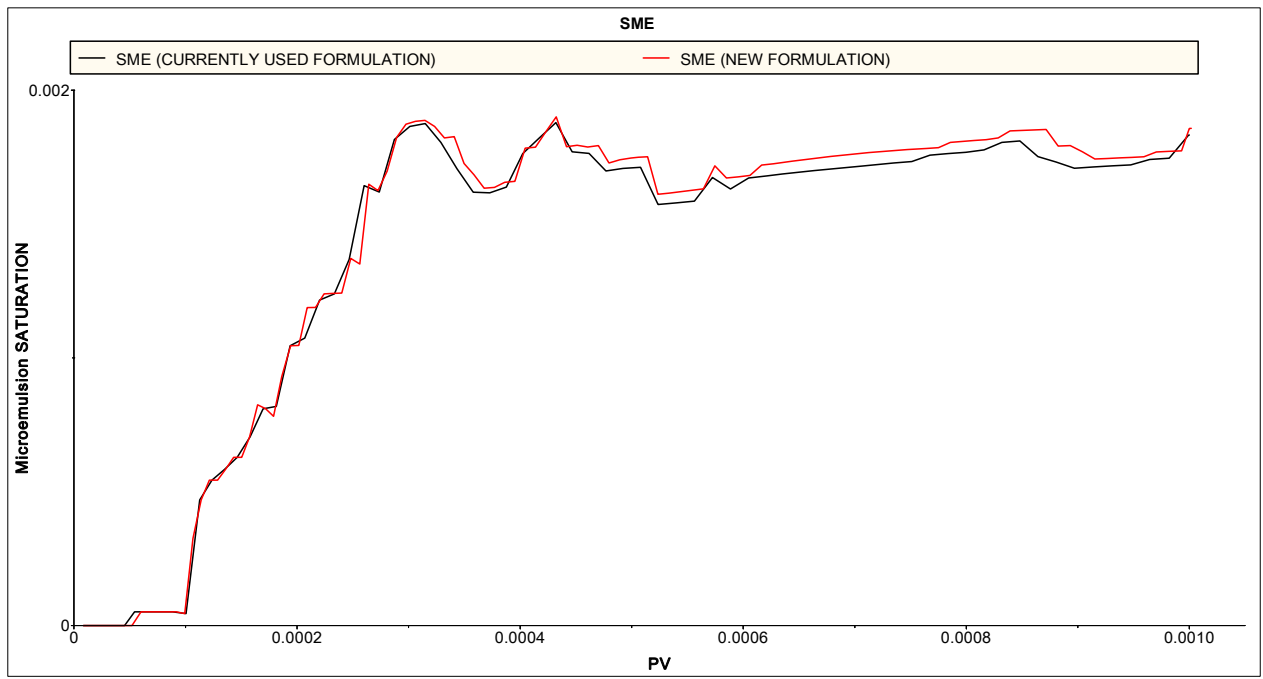

**Figure 5** Microemulsion saturation vs. injected pore volume

As it can be seen from the above theoretical analysis, the new and currently used formulations are very much the same for the case of neglected volume fraction of the components adsorbed onto the solid surface (i.e., by setting $\sum_{i=1}^{n_{cv}} \hat{c}_i = 0$). In this work, through its application to the above-mentioned numerical experiment and comparisons with existing chemical compositional model's numerical results for the case where adsorption effects on the transport of a component are negligible, the newly developed formulation has proven to be practical and reliable.

We think that there is no need in demonstrating here the porosity change effect comparing the results for the case when change in porosity due to the adsorption influences the process significantly, because the mathematical proof provided above is much convincing than a numerical proof. The implementation of a sequential solution approach and comparison in terms of speed and accuracy with the IMPEC approach is scheduled for the future.

# Conclusion

In the scope of this research work, a new mathematical formulation of the chemical compositional reservoir flow equations for the sequential simulation has been developed. The mathematical formulation developed in the scope of this work is extended from the existing model formulation for use in chemical flooding studies. During the process of this research, it was revealed that the approach used in the existing chemical compositional model estimates the adsorption effect on the transport of a component reasonably well but it does not satisfy the species conservation equation. Since the energy conservation equation in the currently used chemical compositional model does not take into account any change in the porosity due to adsorption, the approach leads to inconsistency between the overall compositional balance equations and the energy conservation equation by violating conservation of energy. In the present work we introduce an approach to model the reduction in pore volume due to adsorption that satisfies the conservation laws for mass and energy. One can be convinced of the validity of this conclusion by comparing the energy conservation equation and the reaction rate terms of the species conservation equations in our formulation with those of the currently used chemical compositional model formulation. The mathematical formulation of species conservation equations suggested in this work does not require overall change in the currently used algorithm and makes it possible to apply a sequential solution approach to solve each of these equations separately and implicitly. A comparison with UTCHEM simulator has been performed for the case when change in porosity due to the adsorption does not influence the process significantly. Comparative studies show that the results obtained from IMPEC implementation of the newly proposed formulation are in a good agreement with that of UTCHEM simulator. The implementation of a sequential solution approach for chemical compositional reservoir simulation based on the formulation described in this paper is scheduled for the future.

# Acknowledgements

This research was sponsored and supported by Kazakhstan's Bolashak International Scholarship Program, the Research Institute of Mathematics and Mechanics at Al-Farabi Kazakh National University under grant No. 1735/GF4, and the Republican State-Owned Enterprise "Institute of Information and Computational Technologies" of the Science Committee of the Ministry of Education and Science of the Republic of Kazakhstan under grant No. 0128/GF4.
The first author sincerely thanks Professor Kamy Sepehrnoori and other colleagues for their generous hospitality at The University of Texas at Austin, where this research was started. Professor Sepehrnoori and his research team provided insight and expertise that greatly assisted the research. Thanks are also due to Sciencesoft Ltd for providing the software used in this work.